\documentclass[prb,aps,amssym,nofootinbib,floatfix,reprint,notitlepage]{revtex4-2} 

\usepackage{xcolor}
\usepackage{amsmath,amssymb,bbold,bm}
\usepackage{graphicx}
\usepackage[normalem]{ulem}
\usepackage{cancel}
\usepackage{comment}

\usepackage[pdfpagelabels,colorlinks=true,breaklinks=true,linktocpage=true,linkcolor=red,urlcolor=blue,citecolor=blue,%
]{hyperref}

\newcommand{\be}{\begin{equation}}
\newcommand{\ben}{\begin{equation*}}
\newcommand{\ee}{\end{equation}}
\newcommand{\een}{\end{equation*}}
\newcommand{\bs}{\begin{split}}
\newcommand{\es}{\end{split}}
\newcommand{\bmx}{\begin{array}}
\newcommand{\emx}{\end{array}}

\newcommand{\bea}{\begin{eqnarray}}
\newcommand{\bean}{\begin{eqnarray*}}
\newcommand{\eea}{\end{eqnarray}}
\newcommand{\eean}{\end{eqnarray*}}
\newcommand{\dg}{^{\dagger}}
\newcommand{\dn}{^{\vphantom{\dagger}}}

\newcommand{\ua}{\uparrow}
\newcommand{\da}{\downarrow}
\newcommand{\Ua}{\Uparrow}
\newcommand{\Da}{\Downarrow}

\newcommand{\sgn}[1]{{\rm sign}{#1}}
\newcommand{\pref}[1]{(\ref{#1})}

\newcommand{\abs}[1]{\left\vert #1 \right\vert}
\newcommand{\bra}[1]{\left\langle #1 \right\vert}
\newcommand{\ket}[1]{\left\vert #1\right\rangle}
\newcommand{\braket}[1]{\left\langle #1\right\rangle}

\newcommand{\mat}[1]{\left(\bmx{cc}#1\emx\right)}
\newcommand{\matc}[2]{\left(\bmx{#1}#2\emx\right)}

\newcommand{\bw}[1]{\begin{widetext}}
\newcommand{\ew}[1]{\end{widetext}}

\newcommand{\gray}[1]{}

\newcommand{\nothing}[1]{}

\begin{document}

\title{Spin fractionalization in a Kondo-lattice superconductor heterostructure}
\author{Ethan Huecker}
\author{Yashar Komijani\,$^{*}$}
 \affiliation{Department of Physics, University of Cincinnati, Cincinnati, Ohio, 45221, USA}
\date{\today}
\begin{abstract}
Kondo lattices are one of the classic models of strongly correlated systems where despite a long history, a full understanding of the excitation spectra is still not available. Here we propose that recent progress in engineering heterostructures can be leveraged to gain insight into and even tune this spectra. We use a strong Kondo coupling expansion to study spin-1 excitations of a Kondo lattice in both one and two-dimension to see whether or not paramagnons in a Kondo insulator fractionalize into spin-1/2 excitations. We show that while paramagnons are stable in the strong Kondo coupling limit, presence of sufficient proximity-induced superconducting pairing can favor fractionalization. Our results can be checked using a neutron scattering study of Kondo lattice heterostructures and paves the way toward engineering strongly correlated electronic systems.
\end{abstract}
\maketitle

\section{Introduction}
The Kondo lattice (KL) model has emerged as a powerful theoretical framework for understanding the interplay between itinerant conduction electrons and localized magnetic moments in strongly correlated materials \cite{Hewson1993,Coleman2016}. The one-dimensional version, in particular, has been studied by a number of analytical and numerical techniques, including Monte Carlo \cite{Fye90,Troyer93,Raczkowski19,Danu2021}, density matrix renormalization group (DMRG) \cite{Yu93,Sikkema97,McCulloch99,Shibata99,Smerat09,Peters12,Khait5140}, bosonization \cite{Tsvelik94,Zachar96,Tsvelik19}, strong-coupling expansion \cite{Sigrist97,Trebst06}, and exact diagonalization \cite{Basylko08}. Furthermore, renormalization and Monte-Carlo methods have also been used to examine the p-wave version of the 1D topological Kondo lattice model, which exhibits topological end-states \cite{Alexandrov2014,Lobos15,Zhong2017,Zhong2018}.

The collected wisdom using these approaches is that in the particle-hole symmetric limit, the model realizes an insulator, featuring a unique ground state with short-range entanglements which can be smoothly evolved into a product state of local Kondo singlets. Upon doping, the model realizes a heavy Luttinger liquid which transitions into a ferromagnet in the strong coupling limit \cite{Shibata99}. 

Despite this, the spectrum of low-energy excitations of the system appears to hold exotic features related to spin fractionalization \cite{Chen2023}, within which charge-neutral spin-1 paramagnons may fractionalize into charged excitations that transform as spin-1/2 representations of the SU(2) group. This phenomena can be viewed as a deconfining transition in they theory composed of coupled gauge and matter fields \cite{Polyakov}. While at the strong coupling the two-particle excitations are confined into a paramagnon, it is suggested in \cite{Chen2023} that in the weak Kondo coupling regime, the paramagnon might be deconfined. This possibility has been raised in connection to the numerical results from 1D matrix product states and is expected to be even more relevant in higher dimensions. 

In this paper, however, we discuss whether or not the deconfinement can be influenced within the strong Kondo coupling limit. We study the possibility of using spectroscopic experiments on engineered heterostructures to not only study the excited states but to also induce transitions in the spectrum. Linear response probes are regularly used to study the properties of the ground states in quantum materials. Since such techniques always probe a transition between the ground state and low-lying excitations, a re-arrangement in the manifold of low-energy excited states can strongly influence the result. Therefore, the aforementioned transitions qualitatively affect the linear response results.

\subsection{Kondo lattice model}
The goal is to study the standard Kondo lattice model which is described by the Hamiltonian
\be
H_{\rm KL}=J_K\sum_{\bf j}\Vec{S}_{\bf j}\cdot c\dg_{\bf j}\Vec{\sigma}c\dn_{\bf j}-t\sum_{\braket{\bf ij},\sigma}\left(c\dg_{{\bf i},\sigma}c\dn_{{\bf j},\sigma}+\text{h.c.}\right).\label{eq_HKL}
\ee
Here, $\braket{\bf ij}$ refers to nearest neighboring sites and $\sigma=\ua,\da$ is the spin. We use bold face ${\bf i},{\bf j}$ to refer to lattice sites on a one-dimensional or two-dimensional lattice. The Kondo term proportional to $J_K$ represents the antiferromagnetic coupling between localized spins (magnetic moments) $\vec{S}_j$ and delocalized (conduction) electrons, while the second term describes the hopping of electrons on a 1D or 2D square lattice. When the spin-index is suppressed, the annihilation operator $c_{\bf n}$ is regarded as an spinor.

\subsection{Strong coupling limit}

Strong coupling expansion has been applied to the Kondo lattice model before \cite{Sigrist97,Trebst06}. Recent work \cite{Chen2023} has revisited the 1D Kondo insulator using a combination of tensor network, strong coupling expansion, and mean-field analysis. It was shown that in contrast to previous results \cite{Trebst06}, for $J_K/t\sim 2$ the two-particle excitation spectrum of the Kondo insulator is dominated by a paramagnon at lowest energies. At strong coupling this has a simple interpretation as an attractive interaction that binds the doublon and holon into a bound state.

To see this, we use a strong Kondo coupling approximation, $J_K/t\gg 1$, under which the first term in \pref{eq_HKL} is the dominating term, favoring singlets at every site. In this limit, the ground state is the product state
\be
\ket{\Omega}=\prod_{\bf n}{\ket{K_n}, \qquad \ket{K_n}=}\frac{\ket{\Ua_{\bf n}\da_{\bf n}}-\ket{\Da_{\bf n}\ua_{\bf n}}}{\sqrt 2},
\ee
with energy $E_0=-(3J_K/2){\cal N}$, where ${\cal N}$ is the number of sites. {The charge-e, spin-1/2 single-particle excitations on top of this ground state are known as doublons $c\dg_{{\bf n}\sigma}\ket{\Omega}$ and holons $c\dn_{{\bf n}\sigma}\ket{\Omega}$ \cite{Trebst06,Chen2023} with an energy of $E_1=E_0+3J_K/2$.} Our primary interests are the spin-1 excitations which in the large-$J_K$ limit correspond to a simultaneous creation of a doublon and holon on the same site \cite{Chen2023}.
As long as the ground state is a Kondo singlet, we can write
\be
\left(\Vec{S}_{\bf n}+c^\dagger_{\bf n}\frac{\Vec{\sigma}}{2}c\dn_{\bf n}\right)\ket{\Omega}=0,\label{eq_singlet}
\ee
which indicates that a spin-flip excitation with the Kondo energy $E_t=E_0+2J_K$ is equivalent to a two-particle (charge-2e) excitation at the same site. A doublon-holon (DH) state can be written generally as the superposition
\be
\ket{\rm DH}=\sum_{\bf n_1n_2}\psi_{\rm DH}({\bf n_1},{\bf n_2})c\dg_{{\bf n_1}\ua}c\dn_{{\bf n_2}\da}\ket{\Omega}\label{eq_DH}.
\ee
If not on the same site, a doublon-holon state has the Kondo energy $E_2=E_0+3J_K$. This is schematically shown in Fig.\,\ref{fig1}(a). A spin-flip can in principle fractionalize into an independent doublon and holon, propagating through the lattice and later recombining into a spin-excitation. However since $E_t<E_2$, a paramagnon is energetically stable and does not fractionalize. This can be interpreted in terms of presence of an attractive interaction $V=-J_K<0$ that binds the doublon and holon into a paramagnon.

\subsection{Model - Additional terms}
\begin{figure}[t]
\vspace{1cm}
    \includegraphics[width=\linewidth]{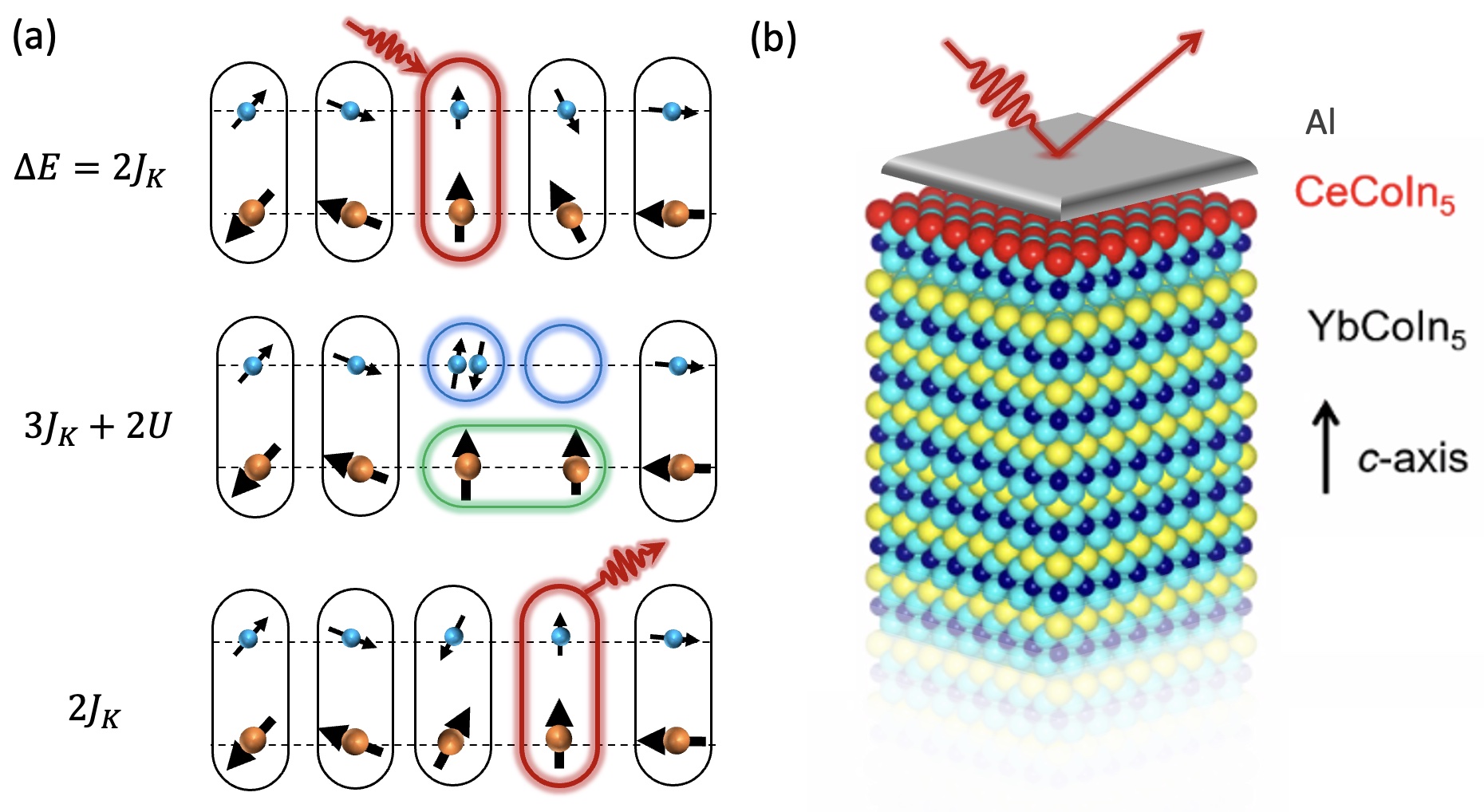}
    \caption{\small  {(a) Kondo lattice model in 1D. A spin-1 spin-flip excitation can propagate through the lattice either as a whole or break up into fractionalized spin-1/2 doublons and holons in between and then recombine again. (b) A Kondo lattice - superconductor heterostructure can be used to engineer the spin dynamics of the system, as probed by X-ray scattering.}}    \label{fig1}
\end{figure}

The stability of the paramagnon is not set in stone. For example, adding a local interaction to the Hamiltonian
\be
H_U=U\sum_j\Big(\sum_\sigma c\dg_{j\sigma}c\dn_{j\sigma}-1\Big)^2,
\ee
changes the energies of the doublon/holon to $E_1{=} E_0+3J_K/2+U$ and $E_2{=} E_0+3J_K+2U$ without affecting the paramagnon energy [Fig.\,\ref{fig1}(a)]. The attractive interaction between the holon and doublon is modified to $V{=} -J_K-2U$. Therefore for an attractive $U<-J_K/2$, it is energetically favorable for the spin to fractionalize.

The negative sign of $U$ makes it impossible to verify this prediction. In order to tune the fractionalization in an actual experiment, in this paper we consider adding a singlet s-wave pairing between conduction electrons
\be
H_\Delta=\Delta\sum_j\left(c^\dagger_{j,\uparrow}c^\dagger_{j,\downarrow}+\text{h.c.}\right)\label{eq6}
\ee
to the Hamiltonian, where we assumed $\Delta$ to be real. Such a term can be induced by depositing a thin layer of Al on the epitaxially grown CeCoIn$_5$ as shown in Fig.\,\ref{fig1}(b). Alternatively, it can be produced by including a superconducting layer \cite{Li2023} in twisted multilayer graphene heterostructures where the hybridization of an itinerant band with a localized band is already established \cite{Padmanabhan2022}.

Our primary goal is to examine whether superconducting proximity, represented by $H_\Delta$, can by itself favor fractionalized spin excitations. At first sight it seems that the spin-singlet superconductivity only acts in the charge sector and is unlikely to affect the spin sector. However, superconductivity can transform spin-1 doublon-holon excitations of the Kondo lattice into either doublon-doublon or holon-holon excitations. Since the latter two are not confined the hybridization may induce deconfinement.

In the next section we introduce the method we use to study the problem and provide technical details of our calculations. The readers interested in results can skip to section \ref{sec:results} where the results are presented. The paper is ended with a conclusion and some open questions.

\section{Method}\label{sec:method}

We consider a model that in its most general form includes all the terms introduced, $H=H_{\rm KL}+H_U+H_\Delta$. To tackle this problem we use a strong Kondo coupling approximation, $J_K/t\gg 1$, according to which $H_K$ is the dominating term. For concreteness, we assume a periodic boundary condition with $L$ sites in every direction.

\subsection{Schr\"odinger Equations}
\bw

Here, we focus on two-particle charge-2e and spin-1 excitations, which using Eq.\,\pref{eq_singlet} is most generally, written as 
\be
\ket{F}=\sum_{\bf n_1n_2}\Big[\psi_{\rm DD}({\bf n_1},{\bf n_2})c\dg_{{\bf n_1}\ua}c\dg_{{\bf n_2}\ua}+\psi_{\rm DH}({\bf n_1},{\bf n_2})c\dg_{{\bf n_1}\ua}c\dn_{{\bf n_2}\da}+\psi_{\rm HH}({\bf n_1},{\bf n_2})c\dn_{{\bf n_1}\da}c\dn_{{\bf n_2}\da}\Big]\ket{\Omega}.\label{eq_F}
\ee
The three terms on the right are doublon-doublon $\ket{\rm DD}$, doublon-holon $\ket{\rm \rm DH}$, and holon-holon $\ket{\rm HH}$ states respectively.  For the $\ket{\rm DD}$ and $\ket{\rm HH}$ states we use the anti-symmetry of the wavefunction $\psi_{\rm DD/HH}({\bf n_1},{\bf n_2})=-\psi_{\rm DD/HH}({\bf n_2},{\bf n_1})$ to limit ourselves to $n_1>n_2$ case. By applying $H\ket{F}=E\ket{F}$ and matching the operators from both sides we obtain the single-particle Schr\"odinger equations
\bea
&-&\frac{t}{2}\sum_{\hat\delta}[\psi_{\rm DD}({\bf n_1}+\hat \delta,{\bf n_2})+\psi_{\rm DD}({\bf n_1},{\bf n_2}+\hat \delta)]+\Delta\psi_{\rm DH}({\bf n_1},{\bf n_2})-\Delta\psi_{\rm DH}({\bf n_2},{\bf n_1})=(E-E_2)\psi_{\rm DD}({\bf n_1},{\bf n_2})\label{psidd}\\
&-&\frac{t}{2}\sum_{\hat\delta}[\psi_{\rm DH}({\bf n_1}+\hat \delta,{\bf n_2})-\psi_{\rm DH}({\bf n_1},{\bf n_2}+\hat \delta)]+V\delta_{{\bf n_1},{\bf n_2}}\psi_{\rm DH}({\bf n_1},{\bf n_2})\label{psidh}\\
&&\hspace{7cm}+\Delta\psi_{\rm DD}({\bf n_1},{\bf n_2})+\Delta\psi_{\rm HH}({\bf n_1},{\bf n_2})=(E-E_2)\psi_{\rm DH}({\bf n_1},{\bf n_2})\nonumber\\
&+&\frac{t}{2}\sum_{\hat\delta}[\psi_{\rm HH}({\bf n_1}+\hat \delta,{\bf n_2})+\psi_{\rm HH}({\bf n_1},{\bf n_2}+\hat \delta)]+\Delta\psi_{\rm DH}({\bf n_1},{\bf n_2})-\Delta\psi_{\rm DH}({\bf n_2},{\bf n_1})=(E-E_2)\psi_{\rm HH}({\bf n_1},{\bf n_2}).\label{psihh}
\eea

\ew

Here $\hat\delta\in\{\hat x, -\hat x\}$ for the one-dimensional problem and $\hat\delta\in\{\hat x, -\hat x,\hat y,-\hat y\}$ for the two-dimensional problem. {We remind the reader that $\Delta$ is the superconductivity pairing amplitude and $V=-J_K-2U$ is the strength of on-site interaction between doublons and holons. Note that the on-site interaction $V$ does not act on the $\Psi_{DD}$ or $\Psi_{HH}$ because due to Pauli principles these states vanish when the corresponding particles are on the same site.}

\subsection{Making use of the translational invariance}
Translational invariance requires that upon an equal shift of positions ${\bf n_1}$ and ${\bf n_2}$, the wavefunction can only acquire a phase shift. This implies a separation of variables
\be
    \psi({\bf n_1},{\bf n_2})=e^{i{\bar{\bf k}}\cdot{\bar{\bf n}}}\phi(\Delta{\bf n};{\bar k}),\label{eq_wf}
\ee
where $\bar{\bf n}=({\bf n_1}+{\bf n_2})/2$ and $\Delta{\bf n}={\bf n_1}-{\bf n_2}$ and $\bar{\bf k}$ is the total momentum of the two-particle complex. It should be clear from the context if $\Delta$ refers to the amplitude of pairing or the difference in positions. This reduces the dimension of the Schr\"odinger equations from $L^2$ to $L$ dimensional matrices, written as
\bw

\bea
&-&t\sum_{\vert{\hat\delta}\vert}\cos(\bar k_{\hat\delta}/2)[\phi_{\rm DD}(\Delta{\bf n}+\hat \delta)+\phi_{\rm DD}(\Delta{\bf n}-\hat \delta)]+\Delta\phi_{\rm DH}(\Delta{\bf n})-\Delta\phi_{\rm DH}(-\Delta{\bf n})=(E-E_2)\phi_{\rm DD}(\Delta{\bf n})\nonumber\\
&-&it\sum_{\vert{\hat\delta}\vert}\sin(\bar k_{\hat\delta}/2)[\phi_{\rm DH}(\Delta{\bf n}+\hat \delta)-\phi_{\rm DH}(\Delta{\bf n}-\hat \delta)]+V\delta_{\Delta{\bf n},{\bf 0}}\phi_{\rm DH}(\Delta{\bf n})
+\Delta\phi_{\rm DD}(\Delta{\bf n})+\Delta\phi_{\rm HH}(\Delta{\bf n})=(E-E_2)\phi_{\rm DH}(\Delta{\bf n})\nonumber\\
&+&t\sum_{\vert{\hat\delta}\vert}\cos(\bar k_{\hat\delta}/2)[\phi_{\rm HH}(\Delta{\bf n}+\hat \delta)+\phi_{\rm HH}(\Delta{\bf n}-\hat \delta)]+\Delta\phi_{\rm DH}(\Delta{\bf n})-\Delta\phi_{\rm DH}(-\Delta{\bf n})=(E-E_2)\phi_{\rm HH}(\Delta{\bf n})\label{eq_Phi}
\eea
where $\vert{\hat\delta}\vert=\hat x$ in 1D or $\hat x,\hat y$ in 2D.

\ew

\subsection{Boundary conditions: one-dimensional case}
We assume a periodic boundary condition, i.e. the system is on a ring in the one-dimensional case. The boundary condition on the DH state is
\be
\hspace{-.2cm}\psi_{\rm DH}({ n_1}+L,{ n_2})=\psi_{\rm DH}({ n_1},{ n_2}+L)=\psi_{\rm DH}({ n_1},{ n_2}),\label{eq_bc1}
\ee
and can be translated to the boundary conditions on $\bar n$ and $\Delta n$. For $\bar n$ we simply have
\be
e^{i\bar{k} L}=1,\qquad \bar{k}=\frac{2\pi}{L}m, \quad m=0\cdots (L-1).
\ee
Imposing the periodic boundary condition on $\Delta n$ in the DH sector is more subtle. With an open boundary condition we would have $\Delta n=-(L-1)\cdots(L-1)$, but in the presence of a periodic boundary condition there is a redundancy. It is helpful to introduce an auxiliary wavefunction $\tilde\phi_{\rm DH}(\Delta n)$, in terms of which $\psi_{\rm DH}(n_1,n_2)=e^{i\bar kn_1}\tilde\phi_{\rm DH}(\Delta n)$, that treats the position of the doublon $n_1$ rather than average position $\bar{n}$ as the reference of coordinate{, i.e. the doublon is at zero and the holon is at $\Delta n$. It is clear that the two wavefunctions are related by $\phi_{\rm DH}(\Delta n)=e^{i\bar k\Delta n/2}\tilde\phi_{\rm DH}(\Delta n)$, for which the wavefunction $\tilde\phi_{\rm DH}(\Delta n)$ has a trivial boundary condition. We are free to label any of the $L$ sites the holon occupies by identifying the end points, with one possible choice as $\lceil{1-L/2}\rceil\le \Delta n\le \lceil{L/2}\rceil$. Identifying the two ends leads to  $\tilde\phi_{\rm DH}(\lceil{L/2}+1\rceil)=\tilde\phi_{\rm DH}(\lceil{1-L/2}\rceil)$ and $\tilde\phi_{\rm DH}(\lceil{-L/2}\rceil)=\tilde\phi_{\rm DH}(\lceil{L/2}\rceil)$. Translating the latter condition to $\phi_{\rm DH}(\Delta n)$ we find}
\be
\phi_{\rm DH}(\lceil{L/2}\rceil)=e^{i\bar k L/2}\phi_{\rm DH}(\lceil{-L/2}\rceil).\label{eq_bc2}
\ee
This result has a simple interpretation. As the two particles move away from each other on a ring, $\Delta n$ initially grows, but at some point when they are at diametrically opposite sides the center of mass moves from one side of the circle to the other. The wavefunctions are equivalent up to a change in the center of mass, which leads to Eq.\,\pref{eq_bc2}. This also follows more directly from the substitution of the wavefunction in Eq.\,\pref{eq_wf} into Eq.\,\pref{eq_bc1}.

A more useful choice is $\Delta n=0\cdots (L-1)$ to conclude that $\tilde\phi_{\rm DH}(L)=\tilde\phi_{\rm DH}(0)$ and $\tilde\phi_{\rm DH}(-1)=\tilde\phi_{\rm DH}(L-1)$. More generally, we have that $\tilde\phi_{\rm DH}(-\abs{\Delta n})=\tilde\phi_{\rm DH}(L-\abs{\Delta n})$. In terms of the $\phi_{\rm DH}(\Delta n)$ wavefunction this means
\be
\phi_{\rm DH}(L)=e^{i\bar k L/2}\phi_{\rm DH}(0),\label{eq_bc3}
\ee
or equivalently $ \phi_{\rm DH}(-1)=e^{-i\bar k L/2}\phi_{\rm DH}(L-1)$. 

For DD and HH states, the boundary condition is simpler. Due to the anti-symmetry of the wavefunction we can focus on $n_1\ge n_2$, or equivalently $0\le\Delta n\le L$. Furthermore, they cannot occupy the same site. Therefore, the boundary condition on $\Delta n$ part is that of open boundary condition
\be
\phi_{\rm DD/HH}(L)=\phi_{\rm DD/HH}(0)=0.
\ee 
\begin{figure}[tp]
\includegraphics[width=\linewidth]{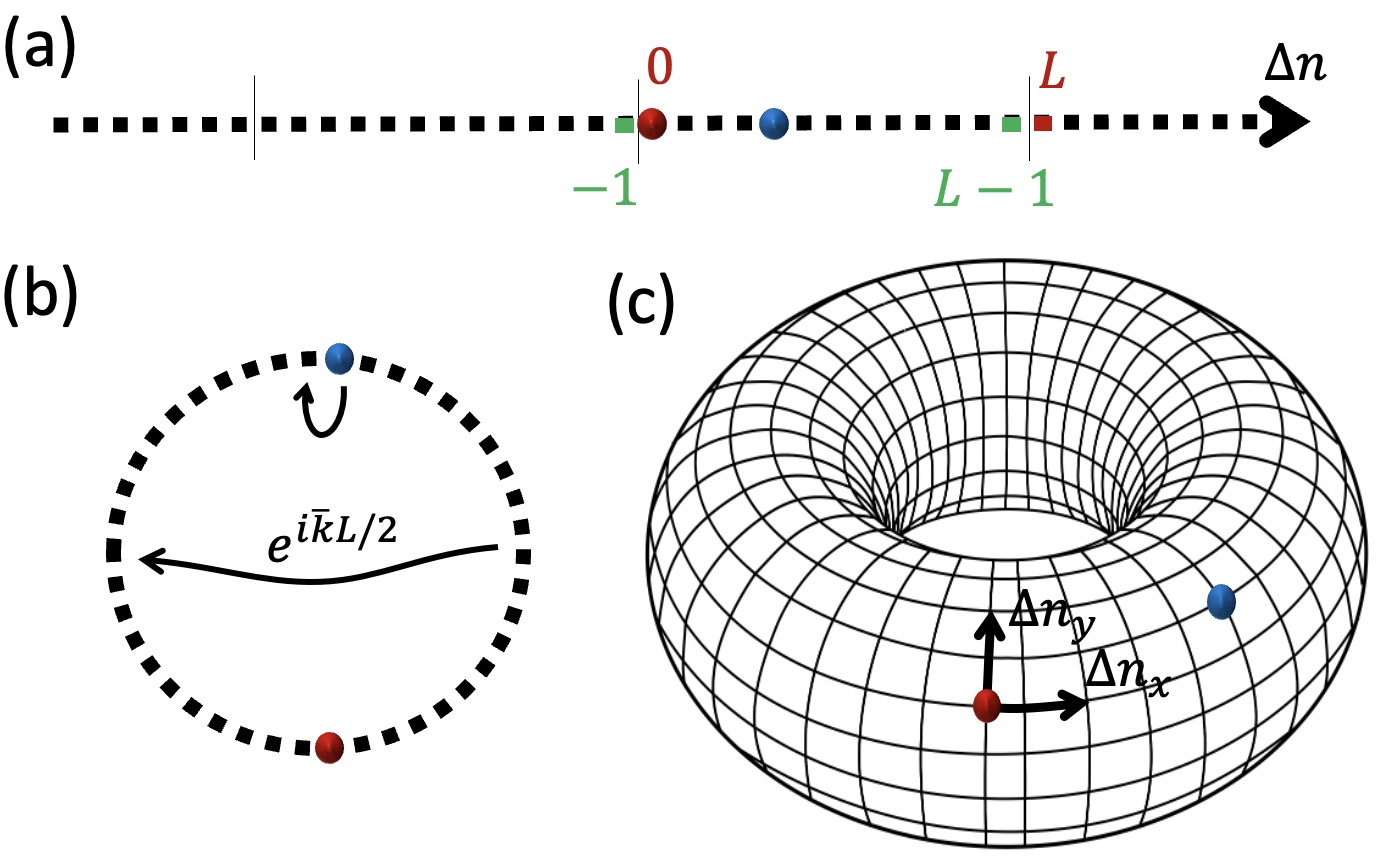}
\caption{\small (a) The periodic boundary condition in 1D can be imposed by identifying the two points of any interval, while keeping one particle (red dot) at $\Delta n=0$. (b) As the particles get farther apart, at some point the center mass coordinate $\bar n=(n_1+n_2)/2$ moves from one side to the other, resulting in a phase shifted boundary condition. (c) The same approach in 2D, the periodic boundary condition can be imposed by fixing either one of the particles at $\Delta{\bf n}={\bf 0}$.}\label{fig2}
\end{figure}
\subsection{Boundary conditions: two-dimensional case}
Again we assume a periodic boundary condition which in 2D means the system lives on a torus. Without losing generality, let us assume $L_x=L_y=L$. The boundary condition Eq.\,\pref{eq_bc1} becomes
\be
\psi_{\rm DH}({\bf n_1}+L\hat\delta,{\bf n_2})=\psi_{\rm DH}({\bf n_1},{\bf n_2}+L\hat\delta)=\psi_{\rm DH}({\bf n_1},{\bf n_2}),\nonumber
\ee
which can be translated to the boundary conditions on $\bar n$ and $\Delta n$. We find that
\be
e^{\bar{\bf k}\cdot L\hat{\bf \delta}}=1,\qquad \bar{\bf k}=\frac{2\pi}{L}{\bf m},
\ee
and
\be
\phi(L\hat x+y\hat y)=e^{\frac{i}{2}\bar k_xL}\phi(y\hat y), \quad \phi(x\hat x+L\hat y)=e^{\frac{i}{2}\bar k_yL}\phi(x\hat x).\nonumber
\ee
For the DD and HH states, it is again helpful to pick one of the particles as the center of coordinate and define $\psi({\bf n_1},{\bf n_2})=e^{i\bar{\bf k}\cdot{\bf n_1}}\tilde\phi(\Delta{\bf n})$. Since the two-particles live on a torus [Fig.\,\pref{fig2}], both $\Delta n_x$ and $\Delta n_y$ can each be chosen to vary over $0\cdots (L-1)$, with the condition that $\Delta{\bf n}={\bf 0}$ (two particles occupying the same site) is eliminated. Note that choice of the center of coordinate is irrelevant since
\bea
\tilde\phi(-\abs{n_x},n_y)&=&\tilde\phi(L-\abs{n_x},n_y),\nonumber\\
\tilde\phi(n_x,-\abs{n_y})&=&\tilde\phi(n_x,L-\abs{n_y}).
\eea
\subsection{Dynamic spin susceptibility}

The dynamic spin susceptibility has the form
\be \chi_S({\bf n},t)=-i\theta(t)\braket{\Omega\Big\vert[S^+_{\bf n}(t),S^-_{\bf 0}({ 0})]\Big\vert\Omega}.\label{eq_suscp}
\ee
This correlation function can be experimentally measured via inelastic X-ray or neutron scattering \cite{Furrer2009}. The $S^-({\bf 0},0)$ operator creates a triplet state at site $\bf n=\bf 0$ at time $t=0$, the triplet propagates through the lattice, and is measured by $S^+({\bf n},t)$ at a later time $t$ on the site $\bf n$ by annihilating the triplet.

Eq.\,\pref{eq_singlet} indicates that as long as the ground state is a Kondo singlet, a spin flip is equivalent to a two-particle (charge-2e) excitation. Therefore $S^+_{\bf n}$ and $S^-_{\bf n}$ in Eq.\,\pref{eq_suscp} can be replaced with $c\dg_{\bf n\ua}c\dn_{\bf n\da}$ and its hermitian conjugate, respectively. These operators act in the two-particle sector, within which the resolution of the identity can be expressed as
\bea
\hat 1&=&\sum_{\bar {\bf k}}\sum_\lambda\ket{F_{\bar{\bf k},\lambda}}\bra{F_{\bar{\bf k},\lambda}},\label{eq_identity}
\eea
where the eigenstates $\ket{F}$ in Eq.\,\pref{eq_F} have DD, DH, and HH sectors. Here, $\lambda$ is the label for the energy eigenstates for a given momentum ${\bf k}$.
Using Eq.\,\pref{eq_identity} and Eq.\,\pref{eq_singlet} into Eq.\,\pref{eq_suscp}, only the DH sector is picked by $c\dg_{\bf n\ua}c\dn_{\bf n\da}$ and the susceptibility becomes

\bw

\bea
    \chi_S({\bf n},t)&=&-i\theta(t)\sum_{\Bar{k},\lambda}\Big[e^{-it(E_{\bf k},\lambda-E_0)}\braket{\Omega\vert c^\dagger_{{\bf n}\uparrow}c_{{\bf n}\downarrow} \vert F_{\Bar{k},\lambda}}\braket{F_{\Bar{k},\lambda} \vert c^\dagger_{{\bf 0}\downarrow}c\dn_{{\bf 0}\uparrow} \vert \Omega}-e^{it(E_{{\bf k},\lambda}-E_0)}\braket{\Omega\vert c^\dagger_{{\bf 0}\downarrow}c\dn_{{\bf 0}\uparrow} \vert F_{\Bar{k},\lambda}}\braket{F_{\Bar{k},\lambda} \vert c^\dagger_{{\bf n}\uparrow}c_{{\bf n}\downarrow} \vert \Omega}\Big]\nonumber\\
    &=&-i\theta(t)\sum_{\Bar{k},\lambda}\Big{[}e^{it(E_0-E_{{\bf k},\lambda})}\psi\dn_{\rm DH,{\bf k},\lambda}({\bf n},{\bf n})\psi^*_{\rm DH,{\bf k},\lambda}({\bf 0},{\bf 0})-e^{it(E_{{\bf k},\lambda}-E_0)}\psi_{\rm DH,{\bf k},\lambda}({\bf 0},{\bf 0})\psi^*_{\rm DH,{\bf k},\lambda}({\bf n},{\bf n})\Big{]}.\label{chichi}
\eea

\ew

After substituting Eq.\,\pref{eq_wf} into Eq.\,\pref{chichi} and performing a Fourier transform, we find that the imaginary part of $\chi_S(q,\omega+i\eta)$ for positive $\omega$ is
\be 
-\chi''_S({\bf q},\omega>0)=\sum_{\lambda}\left|\phi_{\rm DH,{\bf q},\lambda}(\Delta {\bf n}= {\bf 0})\right|^2\delta\left({\omega-\xi_{{\bf q},\lambda}}\right),\label{eq_suscp}
\ee
where $\xi_{{\bf q},\lambda}=E_{{\bf q},\lambda}-E_0$ is the excitation energy. Therefore, the knowledge of energy eigenvalues of the two-particle excitations and the equal-position DH admixture of the wavefunctions is sufficient to determine the dynamic spin susceptibility.

{We should also comment on renormalization of $\ket{\Omega}$ and $\ket{F}$ by the hopping term $t$. The hopping term creates virtual DH pairs that appear as admixture to the product state \cite{Chen2023} but these DH pairs are in singlet sector and as such do not mix with the triplet sector DH. Consequently these virtual singlet DH pairs cost $3J_K$ more and their contributions are suppressed in the strong coupling limit.}

\section{Results}\label{sec:results}
In this section, we use the machinery developed above to study the spin susceptibility of Kondo lattices in the presence of electronic interaction and superconducting proximity.  {In all cases, the Schr\"odinger equations with the corresponding boundary conditions are mapped to eigenvalue problems which are solved to extract the eigenvalues and eigenvectors. The latter two are then used to reconstruct the dynamical spin susceptibility.}

We first start with one dimension, where the analytical solution in absence of superconductivity can provide some insight, then include the superconductivity numerically and finally move on to the two-dimensional case. 

\subsection{1D case without superconductivity}
First, we look at the problem in absence of superconductivity $\Delta=0$. Note that the three equations Eqs.\,(\ref{psidd}-\ref{psihh}) decouple in this limit and we are interested only in the $\ket{\rm DH}$ sector. 

For the time being we will label our wavefunctions with good quantum-numbers $k_1$ and $k_2$, which are the momenta of the doublon and holon respectively. The total momentum $\Bar{k}$ is conserved due to translational invariance of the model, meaning that such $k_1,k_2$ state can be mixed with $k'_1=k_1+q$ and $k'_2=k_2-q$. When the doublon and holon are far apart $|\Delta n|\gg 0$, the potential is absent and the associated energy conservation condition is
\be 
\cos(k_1)-\cos(k_2)=\cos(k_1+q)-\cos(k_2-q),\label{eq_dhgq}
\ee
since doublons and holons have opposite dispersion. This has solutions which fix $k_1'=\pi-k_2$ and $k_2'=\pi-k_1$.

For convenience we choose the related quantum numbers $\Bar{k}=k_1-k_2$ and $\Delta k=(k_1+k_2)/2$, interpreted as the total and relative momentum respectively. Suppressing the DH label and using translational invariance, we can write
\be
    \psi_{\Bar{k},\Delta k}(n_1,n_2)=e^{i\Bar{k}\Bar{n}}\phi_{\Delta k}(\Delta n),
\ee
where $\Bar{n}=(n_1+n_2)/2$ and $\Delta n=n_1-n_2$. With this ansatz, the 2D equation Eq.\,\pref{psidh} becomes the 1D equation
\bea
&&-it\sin(\Bar{k}/2)\left[\phi(\Delta n+1)-\phi(\Delta n-1)\right]\nonumber\\
    &&\hspace{1.5cm}+V\delta_{\Delta n,0}\phi(\Delta n)=(E-E_2)\phi(\Delta n).\label{eq_dhsq}
\eea
For $|\Delta n|\gg 0$, the potential drops out and a plane-wave solution $\phi(\Delta n)\sim e^{i\Delta k\Delta n}$ implies that
\be
    E_{\rm DH}=E_2+2t\sin(\Bar{k}/2)\sin(\Delta k)\label{edh}
\ee
for an appropriate $\Delta k$ is the energy of the $\ket{\rm DH}$ state.

For a general wavefunction, we assume the ansatz
\be
    \phi(\Delta n)=\begin{cases}
    Ae^{i\Delta k\Delta n}+A'e^{i(\pi-\Delta k)\Delta n} & \Delta n<0 \\
    C & \Delta n=0\\
    Be^{i\Delta k\Delta n}+B'e^{i(\pi-\Delta k)\Delta n} & \Delta n>0
    \end{cases}\label{eq_wf1d}
\ee
where $e^{i\Delta k\Delta n}$ and $e^{i(\pi-\Delta k)\Delta n}$ plane waves are mixed, as  a consequence of Eq.\,\pref{eq_dhgq}. We also need to impose the periodic boundary condition. Choosing the interval $\lceil{1-L/2}\rceil\le\Delta n\le \lceil{L/2}\rceil$ makes it clear that this is a scattering problem. However, for calculations it is more convenient to choose $0\le\Delta n\le L$, which allows us to use only the $\Delta n\ge 0$  part of Eq.\,\pref{eq_wf1d}.
We then substitute Eq.\,\pref{eq_wf1d} into Eq.\,\pref{eq_dhsq} for the cases of $\Delta n=-1,0,1$. For $\Delta n=1$, Eq.\,\pref{eq_dhsq} enforces the continuity condition $C=B+B'$, which is evident from Eq.\,\pref{eq_wf1d}. The other two cases can be rewritten in the following matrix equation,

\bw

\be
\mat{
1-e^{ik_2L} & 1-(-1)^Le^{-ik_1L}\\e^{-i\Delta k}(1-e^{ik_2L})-u&\qquad-e^{i\Delta k}[1-(-1)^Le^{-ik_1L}]+u
}
\mat{
        B\\B'
}=0,\label{eq_mat}
\ee

\ew

where  we have defined
\be
u\equiv\frac{V/it}{\sin(\bar{k}/2)}.
\ee
For non-trivial solutions, the determinant of the matrix in Eq.\,\pref{eq_mat} has to vanish, giving rise to the equation
\be
u=-2\cos(\Delta k)\frac{\left(1-(-1)^Le^{-ik_1L}\right)\left(1-e^{ik_2L}\right)}{(-1)^Le^{-ik_1L}-e^{ik_2L}}\label{vguy}
\ee
between $\bar{k}$ and $\Delta k$. The associated eigenvector has components which enforce $B=B'e^{-i(\Bar{k}/2+\Delta k)L}$ such that the wavefunction Eq.\,\pref{eq_wf1d} is determined up to an overall normalization $B'=1/L\sqrt{2}$. Note that the momenta $k_1,k_2$ need not be purely real, as the introduction of an imaginary relative momentum in Eq.\,\pref{eq_wf1d} enforces exponential decay for large separations $|\Delta n|\gg0$. This leads naturally to the formation of the bound state (BS) of a doublon-holon pair, which we interpret as the paramagnon excitation.

To determine the $\Delta k$ of the BS, we note that the energy Eq.\,\pref{edh} of the BS must be a real quantity. This restricts $\sin(\Delta k)$ to be real, such that a general form $\Delta k=\Delta k'+i\Delta k''$ must necessarily be
\be 
    \Delta k=\pm\frac{\pi}{2}+i\Delta k'',
\ee
for a DH bound state. Note that the $\pm$ is determined by $\sgn(E-E_2)$ in Eq.\,\pref{edh}. For this choice of $\Delta k$, the secular equation Eq.\,\pref{vguy} can be recast as
\be 
V=\pm2t\sin(\Bar{k}/2)\sinh(\Delta k''),\label{eqVBS}
\ee
where fixing $\pm$ via Eq.\,\pref{edh} implies that the sign of the physical quantity $V$ fixes the sign of $\Delta k''$.

With the choice of $E<E_2$, to get exponential decay in Eq.\,\pref{eq_wf1d} for $|\Delta n|\gg 0$ it is clear that $B'=0$. Furthermore, a substitution of Eq.\,\pref{eq_wf1d} into Eq.\,\pref{eq_dhsq} forces the remaining coefficients to be equivalent, $B=C$. So, the wavefunction is determined up to an overall normalization. We find that 
\begin{align}
    \phi_{\rm BS}(\Delta n)\propto i^{-\Delta n}e^{-\Delta k''\abs{\Delta n}}\label{eqBS}
\end{align}
is the wavefunction for the doublon/holon BS with 
\be
    E_{\text{BS}}-E_2=-2t\sin(\Bar{k}/2)\cosh(\Delta k''),\label{eqEBS}
\ee
as the BS energy according to Eq.\,\pref{edh}. The exponential decay as $\Delta n$ grows in Eq.\,\pref{eqBS} is the key for confinement. Combining Eqs.\,(\ref{eqVBS},\,\ref{eqEBS}) it can be seen that a BS, with $\Delta k''>0$, has the lowest energy only if $V>0$. Similarly, it can be seen that $V<0$ leads necessarily to a high-energy anti-bound state at $E>E_2$.

\begin{figure}[h!]
        \includegraphics[width=\linewidth]{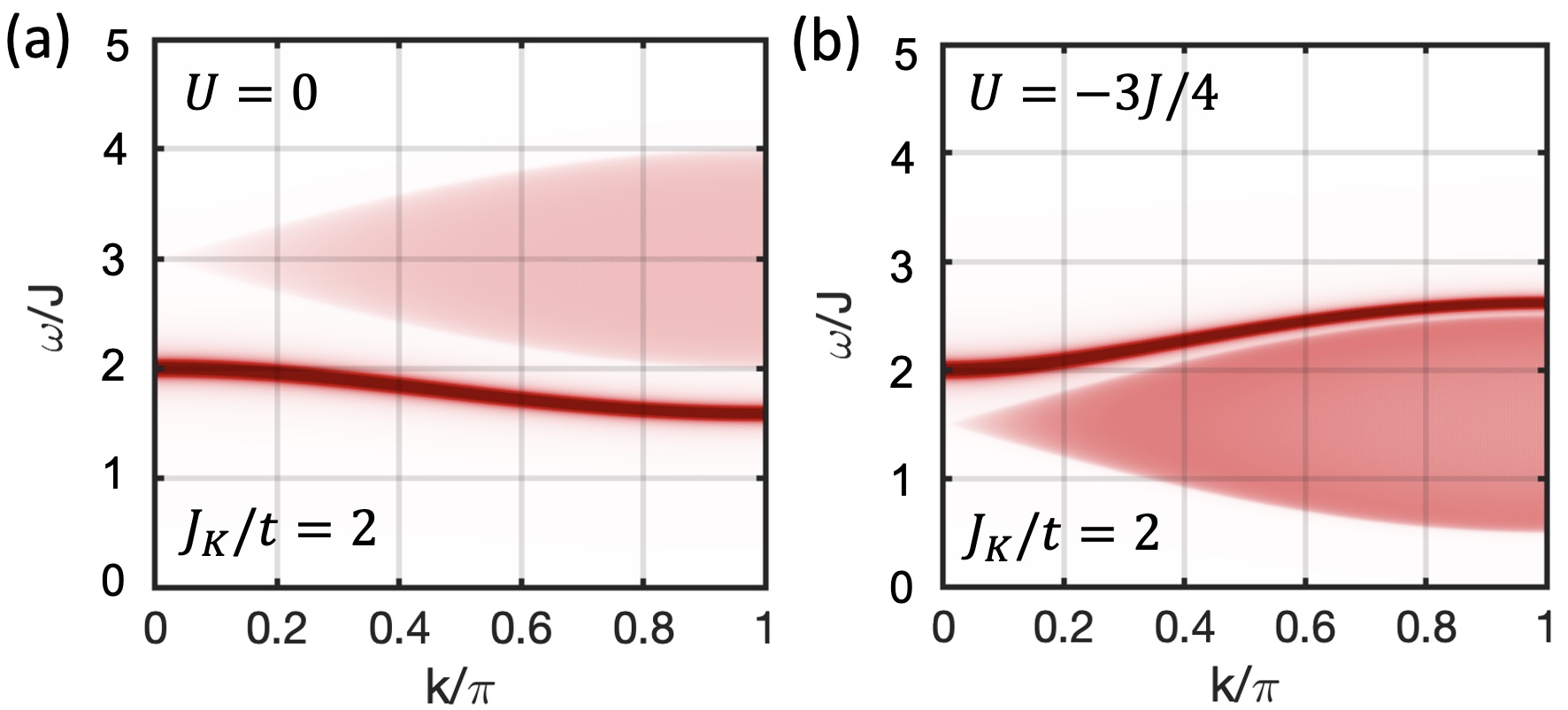}
    \caption{\small  The imaginary part of the dynamical spin susceptibility $-\chi''_S(q,\omega+i\eta)$. (a) At $U=0$ the continuum of excitations form the upper band, while the paramagnon bound state has the lowest energy. (b) For sufficiently attractive interaction among electrons $U<-J/2$, the continuum of excitations is energetically more favorable.}\label{fig3}
\end{figure}

Using Eq.\,\pref{eq_suscp}, the dynamical spin susceptibility becomes
\bea -\chi''_S(q,\omega>0)&=\sum_{\Delta k} \left|C_{q,\Delta k}\right|^2\delta\left({\omega-\xi_{q,\Delta k}}\right),\label{chi}
\eea
where $\xi_{q,\Delta k}$ is defined as
\be
\xi_{q,\Delta k}=E_2-E_0+2t\sin(\Delta k)\sin(q/2).
\ee
The susceptibility is plotted in Fig.\,\pref{fig3} for both $U=0$ and the presence of a finite attractive Coulomb interaction between electrons $U=-3J/4$. Note that for $U<-J/2$ the bound state becomes an anti-bound state and moves to higher energies.

\subsection{{$t=0$ limit and the ground state stability}}
For nonzero $\Delta$, all terms in Eq.\,\pref{eq_Phi} need to be taken into account. It is instructive to first analyze a simplified version of the problem where $t=0$. In this case, the single-doublon or holon states are mixed by the $\Delta$ term.
 This gives rise to new states $\ket{\rm U}$ and $\ket{\rm L}$ of the form
\bea
\ket{\rm U/L}=\frac{\ket{\rm D}\pm\ket{\rm H}}{\sqrt{2}}\otimes\{\ket{\Ua},\ket{\Da}\}.
\eea
These states are denoted as ``upperons" and ``lowerons" by nature of their associated energies $E_{\rm U/L}=E_1\pm\Delta$,
and will be shown to play a key role in explaining certain features of the updated dynamic spin susceptibility. On a two-particle sector, the states would be $\ket{\rm UU}$, $\ket{\rm UL}$, and $\ket{\rm LL}$ with energies $E_2+2\Delta$, $E_2$, and $E_2-2\Delta$ respectively. So, for $t\ll\Delta\ll J$ we expect to see three bands in the two-particle spectrum. The $\ket{\rm DH}$ \emph{orbital}, responsible for the magnetic response of the system, is expressible in terms of $\ket{\rm UU}$, $\ket{\rm UL}$, and $\ket{\rm LL}$ bands.

{Note that Kondo singlets are not affected by the pairing, i.e. $H_\Delta\ket{\Omega}=0$. The fact that for $U=0$ the lowest energies of single-particle and two-particle excitations are modified to $E_{L}-E_0=3J_K/2-\Delta$ and $E_{LL}-E_0=3J_K-2\Delta$ indicates that the $t=0$ ground state composed of product of singlets is stable for $\Delta/J_K<3/2$ and $\ket{\Omega}=\prod_n\ket{K_n}$ neither changes energy nor mixes with other states $H_\Delta\ket{\Omega}=0$ and the initial/final states in Eq.\,\pref{eq_suscp} are still the same $\ket{\Omega}$  defined without pairing. The only intermediate states involved in Eq.\,\pref{eq_suscp} are the two-particle excited states with doublon-holon (DH) pair, whose modification due to pairing is captured via the Schr\"odinger equation. As long as $t\ll J_K$ we expect this assumption to remain valid. In the following we show that even for $\Delta/J_K$ as small as 0.1, there are a critical momenta below which spins are fractionalized.}

\subsection{1D case with superconductivity}
For nonzero hopping, $t\neq0$, we numerically diagonalize the full Hamiltonian Eq.\,\pref{eq_Phi} and use the resulting eigenstates in the computation of the dynamic spin-susceptibility.
\begin{figure}[h!]
        \includegraphics[width=\linewidth]{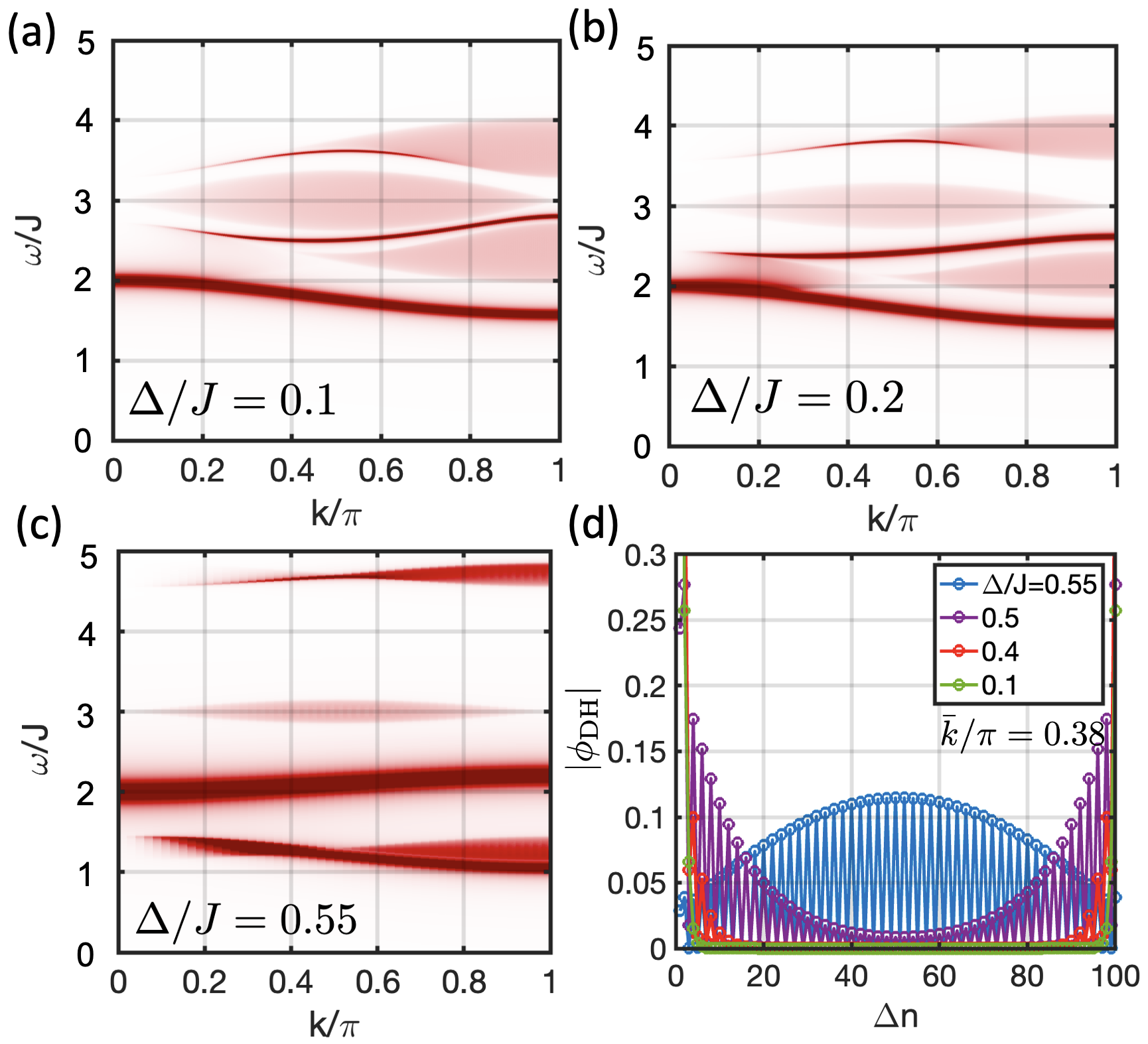}
    \caption{\small  (a-c) The imaginary part of dynamical spin susceptibility $-\chi''_S(q,\omega+i\eta)$ in the presence of superconductivity for $J_K/t=2$ in a $L=100$ site 1D Kondo lattice. As $\Delta$ increases, the upperon and loweron bands separate from the DH continuum, and for sufficient $\Delta$ the loweron merges with the DH bound state, which results in the formation of a new (flipped) bound state. (d) The lowest energy wavefunction in the DH sector for varying $\Delta$. At a $k$-dependent critical pairing $\Delta/J> 0.5$ the boundstate transforms into a free state.}\label{fig4}
\end{figure}
Fig.\,\pref{fig4} shows the effect of superconducting proximity on the dynamic spin susceptibility for various values of $\Delta/J$. The figure shows the formation of upperon/loweron bands that split off the continuum of doublon-holon excitations. Interestingly, new UU, UL and LL bands each inherit the attractive doublon-holon interaction and produce their own (anti-) bound-states.

Fig.\,\ref{fig4}(d) shows the wavefunction $\abs{\phi_{\rm DH}(\Delta n)}$ at a fixed momentum $k/\pi=0.38$. For $\Delta/J<0.5$, the wavefunction is evanescent, indicating the dominance of a bound-state. However, the correlation length increases with increasing $\Delta/J$, and for $\Delta/J>0.5$, the wavefunction transitions to that of a particle-in-a-box, {indicating that the spin-triplet is carried by two spin-1/2 particles rather than a spin-1 bound state}. 

\begin{figure}[h!]
        \includegraphics[width=\linewidth]{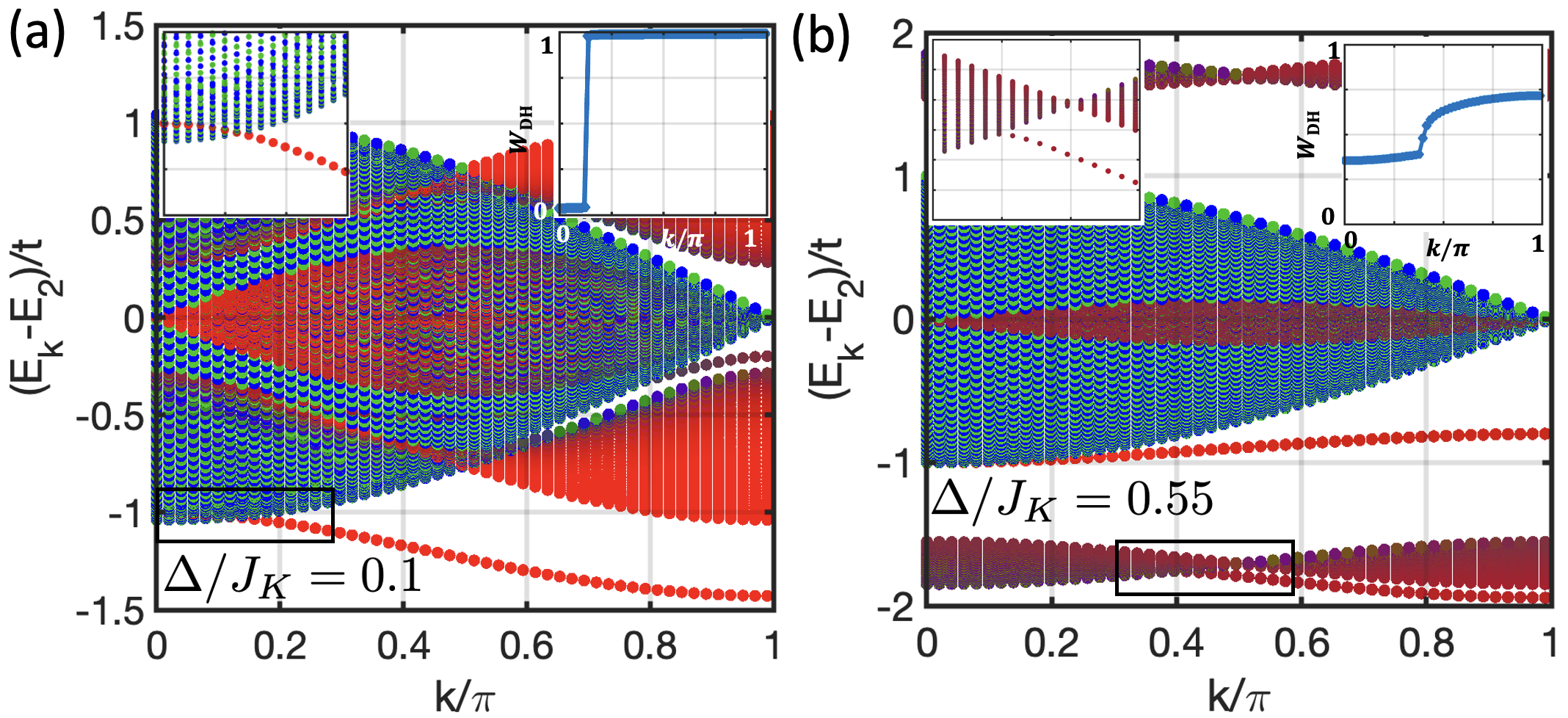}
    \caption{\small  The spectrum of two-particle triplet excitations vs. momentum at (a) $\Delta/J_K=0.1$ and (b) $\Delta/J_K=0.55$. The DH, HH, and DD contents of each eigenstate is shown in red,  green and blue colors respectively. The top left inset shows a zoom into the black rectangle and the top right inset shows the DH weight of the lowest energy state in each case vs. $k$. Note a first-order vs. second-order transition in weight in the two cases.}\label{fig6}
\end{figure}

To gain better insight into this dynamical spin susceptibility, we plot the entire two-particle triplet spectrum in Fig.\,\pref{fig6}. The DH, DD, and HH admixtures of the states are marked by the red, blue and green colors, respectively. The DD/HH states have their largest bandwidths at $k=0$ whereas the DH state has its largest bandwidth at $k=\pi$. Therefore, the BS is hybridized with the DD/HH states for arbitrarily small pairing $\Delta$. For $\Delta/J=0.1$, from the insets of the Fig.\,\ref{fig6}(a) we see that the BS crosses the DD/HH continuum, leading to a sharp transition in the DH character of the lowest triplet state. For $k/\pi<0.15$ the triplet is mostly carried by DD/HH states (and it is deconfined) whereas this character abruptly changes for $k/\pi>0.15$. This is however, not reflected in the spin susceptibility of Fig.\,\ref{fig5}(a) as the latter only probes the DH states. For $\Delta/J=0.55$, the transition moves to larger $k/\pi=0.25$ and the triplet state changes continuously. Since the dynamic susceptibility is only sensitive to the DH admixture with an equal doublon and holon position, this change in character explains the disappearance of the signal at low momenta.

To summarize this section, we observe that an arbitrarily small $\Delta$ is sufficient to induce deconfinement at the lowest momenta. Fast moving magnons are confined, but slow-moving ones are deconfined. Note that for $\bar k>\pi/2$, the BS continues to be the lowest spin-1 excitation of the system.

\subsection{Generalization to 2D}
\begin{figure}[t]
\includegraphics[width=\linewidth]{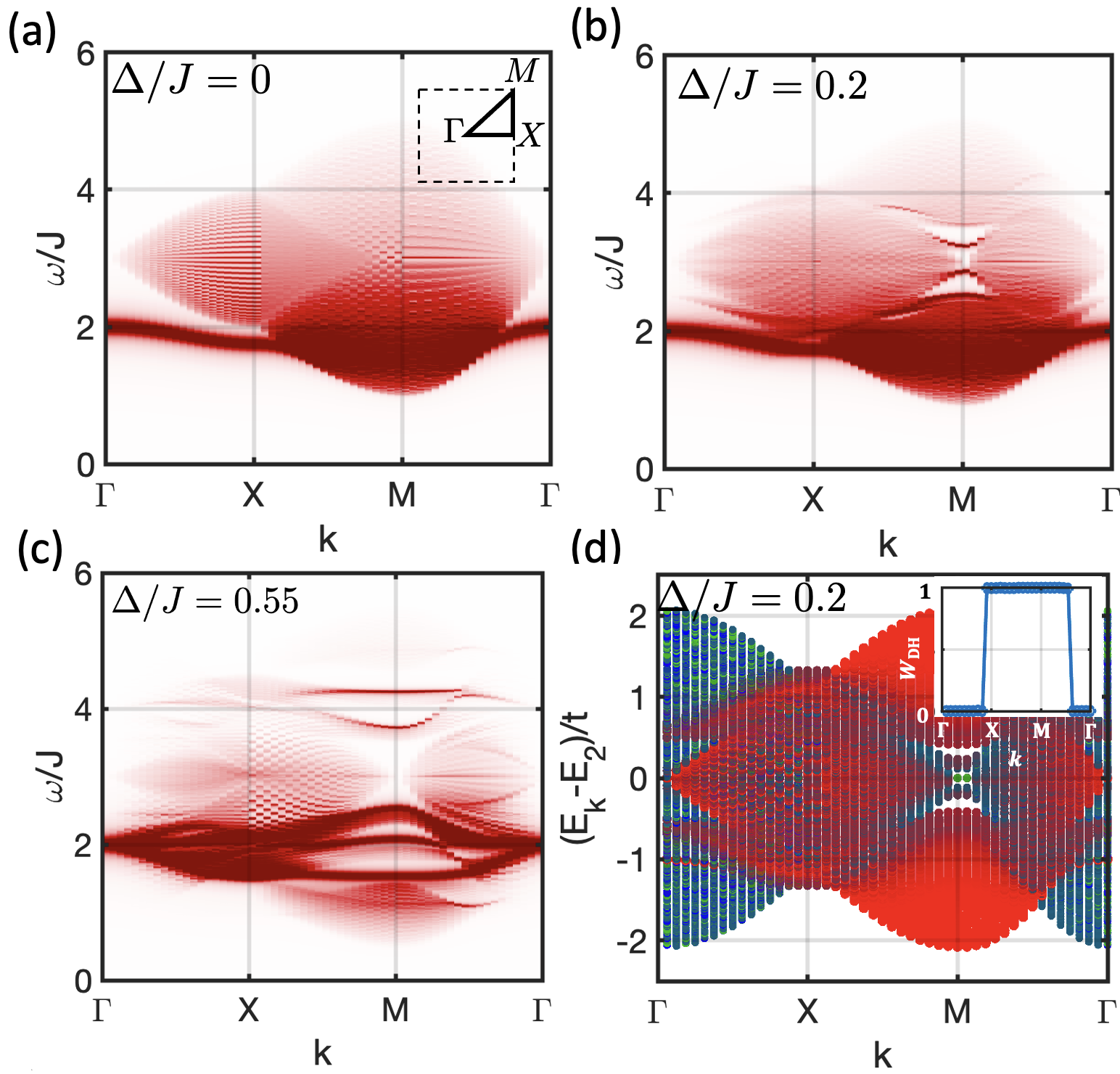}
\caption{\small (a-c) Dynamical spin susceptibility of a 40$\times$40-site 2D Kondo insulator for $J_K/t=2$ and various values of $\Delta/J_K$. The inset of (a) shows the Brillouin zone cut. (d) The two-particle triplet spectrum for $\Delta/J=0.2$. {The inset of (d) shows the DH weight of the lowest energy state.}}\label{fig5}
\end{figure}

The problem can be easily generalized to two-dimension. Solving Eqs.\,\pref{eq_Phi} numerically and using Eq.\pref{eq_suscp} to compute the susceptibility leads to the results shown in Fig.\,\pref{fig5}. This case shares similar features to the 1D version. Namely, at small momenta the spin-excitation is carried by the DD and HH sectors. At higher $\Delta/J$ the spin susceptibility is qualitatively affected, moving the BS to higher energies even in the DH sector.

\section{Conclusion}
In summary, we have studied spin excitations of the 1D and 2D Kondo lattice models at half-filling in the strong Kondo coupling limit. In this limit, the lowest energy excitation is a paramagnon with a continuum band  of fractionalized spin-1/2 excitations at higher energies. 

The significance of the lowest excited states is that typically due to interactions not included explicitly for example electron-phonon coupling, it is expected that the excitation decays into the lowest energetic state with the same quantum numbers. This energetic order can be flipped by a sufficiently attractive Coulomb interaction, without affecting the ground state.

{More practically, however, we showed that a superconducting proximity can cause a deconfining transition and subsequent spin fractionalization.} This is somewhat surprising, as the charge-sector is affecting the spin-sector. In presence of an energy-relaxation mechanism, a small pairing is sufficient to hybridize confined and deconfined sectors and ultimately fractionalize the spins into spin-1/2 doublons and holons at low-momenta. Furthermore, even in absence of energy relaxation, a sufficiently large pairing can lead to deconfinement. Our work opens a window into engineering strongly correlated electronic systems and predictions that can shed light on our understanding of such systems.

The tensor network calculations of \cite{Chen2023} suggest that perhaps in the weak coupling regime $t/J< 1$ the spin fractionalization can take place naturally in the normal Kondo lattice without including other terms considered here. At weak coupling particle-hole excitations (Kondo singlets breaking into virtual doublon-holon pairs) create a plasma of doublons and holons. It is conceivable that such a plasma screens the interaction between the doublon-holon triplet, causing them to be free. Whether such scenario can lead to fractionalization is a fascinating new many-body problem which is outside the scope of the present work and we leave it to future.

\emph{Acknowledgement} - YK acknowledges fruitful discussions with P. Coleman, M. Stoundenmire and J. Chen.

\bibliography{KondoSC}
\end{document}